\documentstyle[12pt,epsfig] {article}
\begin {document}
\begin{center}
\vskip 1.5 truecm
\begin{Large}
$\sigma_L/\sigma_T$ {\bf in the} $\rho^0$ {\bf -meson diffractive
electroproproduction.}\\

\end{Large}
\vspace{.5cm}
M.G.Ryskin and Yu.M.Shabelski \\

\vspace{.5cm}

Petersburg Nuclear Physics Institute, \\
Gatchina, St.Petersburg 188350 Russia \\ \end{center} \vspace{1cm}

\begin{abstract}
The background due to the direct diffractive dissociation of the photon
into the $\pi^+\pi^-$-pair to the "elastic" diffractive $\rho^0$-meson
produc\-tion in electron-proton collisions is calculated. At large $Q^2$ the
interference between resonant and non-resonant $\pi^+ \pi^-$ production
changes the $\sigma^L/\sigma^T$ ratio with the mass of the $2\pi$ (i.e.
$\rho^0$-meson) state.

\end{abstract}
\vspace{3cm}

E-mail: $\;$ RYSKIN@thd.PNPI.SPB.RU  \\

E-mail: $\;$ SHABEL@vxdesy.desy.de \\

\newpage

\section{Introduction}
It was noted many years ago that the form of the $\rho$-meson peak is
distorted by the interference between resonant and non-resonant
$\pi^+\pi^-$ production. For the case of "elastic" $\rho^0$
photoproduction the effect was studied by P.S\"oding in \cite{so} and
S.Drell \cite{dr} (who considered the possibility to produce the pion

beam via the $\gamma\to \pi^+\pi^-$ process).  At high energies the main
(and the only) source of background is the Drell-Hiida-Deck process
\cite{d} (see fig. 1). The incoming photon fluctuates into the pion pair
and then $\pi p$-elastic scattering takes place. Thus the amplitude for
the background may be written in terms of the pion-proton cross section.
Recently the difractive elastic production of $\rho^0$-mesons was
measured at HERA \cite{zeus,zeus1,H1,H11} both for the cases of
photoproduction i.e. $Q^2 = 0$ and of $Q^2 \geq 4$ GeV$^2$ (the so
called deep inelastic scattering, DIS, regime). It was demonstrated
\cite{zeus,H1} that the interference with some non-resonant background
is indeed needed to describe the distribution over the mass - $M$ of
$\pi^+\pi^-$ pair.

In this paper we present the results of calculation of
$R=\sigma^L/\sigma^T$ with correction the numerical error published in
Fig.6 of \cite{RSh}. In Sect. 2 the formulae for the $2\pi$ background
which are valid for the DIS region are presented. The expression differs
slightly from the S\"oding's one as we take into account the pion form
factor and the fact that one pion propagator is off-mass shell. We
consider also the absorbtion correction comming from the diagram where
both pions ( $\pi^+$ and $\pi^-$) directly interact with the target
proton. In Sect. 3 we compute the ratio $R=\sigma^L/\sigma^T$ for a pion
pair production in DIS. At large $Q^2\sim 10\,-\,30$ GeV$^2$ the
background amplitude becomes relatively small, but still not negligible.
It changes the ratio of the longitudinal to transverse $\rho$-meson
production cross section and leads to the decreasing of
$R=\sigma^L/\sigma^T$ value with $M^2$.

\section{Production amplitudes}

The cross section of $\rho^0$ photo- and electroproduction may be written
as:
\begin{equation}
\frac{d\sigma^D}{dM^2dt}\; =\; \int d\Omega
|A_{\rho}+A_{n.r.}|^2 \;,
\end{equation}
where $A_{\rho}$ and $A_{n.r.}$ are the resonant and non-resonant parts of
the production amplitude, $D=L\, ,T$ for longitudinal and transverse photons,
$t = $ -{\bf q}$_t^2$ is the momentum transfered to the proton and
$d\Omega =d\phi dcos(\theta)$, where $\phi$ and $\theta$ are the azimutal
and polar angles between the $\pi^+$ and the proton direction in the
$2\pi$ rest frame.

\subsection{Amplitude for resonant production}

The dynamics of vector meson photo- and electroproduction was discussed
in the framework of QCD in many papers (see, e.g. [9-12]). However here
we will use the simple phenomenological parametrization of the production
amplitude because our main aim is the discussion of the interference
between resonant and non-resonant contributions. So the amplitude for
resonant process $\gamma p \to \rho^0 p$; $\rho^0 \to \pi^+ \pi^-$ reads:
\begin{equation}
A_{\rho}\; =\; \sqrt{\sigma_{\rho}}e^{-b_\rho q^2_t/2}\frac{\sqrt{M_0\Gamma}}
{M^2-M^2_0+iM_0\Gamma}\frac{H^D(\theta ,\phi)}{\sqrt{\pi}} \;.
\end{equation}
To take into account the phase space available for the $\rho\to\pi^+\pi^-$
decay we use the width $\Gamma=\Gamma_0\left(\frac{M^2-4m^2_\pi}
{M^2_0-4m^2_\pi}\right)^{3/2}$ (with $\Gamma_0=151$ MeV and $M_0=768$ MeV
-- its mass); $b_\rho$ is the $t$-slope of the "elastic" $\rho$ production
cross section $\sigma_\rho\equiv d\sigma(\gamma p \to \rho^0 p)/dt$
(at $t=0$) and the functions $H^D(\theta ,\phi),\; D = T, L$ describe the
angular distribution of the pions produced through the $\rho$-meson decay:
\begin{equation}
H^L\; =\; \sqrt{\frac 3{4\pi}}cos\theta \;,
\end{equation}
\begin{equation}
H^T\; =\; \sqrt{\frac 3{8\pi}}sin\theta\cdot e^{\pm i\phi} \;.
\end{equation}
Note that for transverse photons with polarization vector $\vec e$
one has to replace the last factor $e^{\pm i\phi}$ in eq. (4) by the
scalar product $(\vec e\cdot\vec n)$, where $\vec n$ is the unit vector
in the pion transverse momentum direction.

\subsection{Amplitude for non-resonant production}

The amplitude for the non-resonant process $\gamma p \to \pi^+\pi^-p$ is:
\begin{equation}
A_{n.r.}\; =\;\sigma_{\pi p}F_\pi (Q^2)e^{bt/2}\frac{\sqrt{\alpha}}
{\sqrt{16\pi^3}} B^D\sqrt{z(1-z)\left|\frac{dz}{dM^2}\right|
\left(\frac{M^2}4-m^2_\pi\right)|cos\theta|} \;,
\end{equation}
where $b$ is the $t$-slope of the elastic $\pi p$ cross section,
$F_{\pi} (Q^2)$ is the pion electromagnetic form factor
($Q^2=|Q^2_\gamma| > 0$ is the virtuality of the incoming photon),
$\alpha= 1/137$ is the electromagnetic coupling constant and $z$ -- the
photon momentum fraction carried by the $\pi^-$ -meson; $\sigma_{\pi p}$
is the total pion-proton cross section.

The factor $B^D$ is equal to
\begin{equation}
B^D\; =\; \frac{(e^D_\mu\cdot k_{\mu -})f(k'^2_-)}{z(1-z)Q^2+m^2_\pi+k^2_{t-}}
-\frac{(e^D_\mu\cdot k_{\mu +})f(k'^2_+)}{z(1-z)Q^2+m^2_\pi+k^2_{t+}}
\end{equation}
 For longitudinal photons the products $(e^L_\mu\cdot k_{\mu \pm})$ are:
 $(e^L_\mu\cdot k_{\mu -})=z\sqrt{Q^2}$ and
$(e^L_\mu\cdot k_{\mu +})=(1-z)\sqrt{Q^2}$, while for the transverse
photons we may put (after averaging) $e^T_\mu\cdot e^T_\nu =\frac
 12\delta^T_{\mu\nu}$.

Expressions (5) and (6) are the result of straitforward calculation of
the Feynman diagram fig. 1. The first term in (6) comes from the graph
fig. 1 (in which the Pomeron couples to the $\pi^+$) and the second one
reflects the contribution originated by the $\pi^- p$ interaction.  The
negative sign of $\pi^-$ electric charge leads to the minus sign of the
second term. We omit here the phases of the amplitudes. In fact, the
common phase is inessential for the cross section, and we assume that the
relative phase between $A_{\rho}$ and $A_{n.r.}$ is small (equal to zero)
as in both cases the phase is generated by the same 'Pomeron'
\footnote{Better to say -- 'vacuum singularity'.} exchange.

The form factor $f(k'^2)$ is written to account for the virtuality
($k'^2\neq m^2_\pi$) of the t-channel (vertical in fig. 1) pion. As in
fig. 1 we do not deal with pure elastic pion-proton scattering, the
amplitude may be slightly suppressed by the fact that the incoming pion
is off-mass shell. To estimate this suppression we include the form
factor (chosen in the pole form)
\begin{equation}
f(k'^2)=1/(1 + k'^2/m'^2)
\end{equation}
The same pole form was used for $F_\pi (Q^2)=1/(1 + Q^2/m^2_\rho)$.
In the last case the parameter $m_\rho = M_0$ is the mass of the
$\rho$-meson -- the first resonance on the $\rho$-meson (i.e. photon)
Regge trajectory, but the value of $m'$ (in $f(k'^2)$) is expected to be
larger. It should be of the order of mass of the next resonance from
the Regge $\pi$-meson trajectory; i.e. it should be the mass of $\pi
(1300)$ or $b_1 (1235)$. Thus we put $m'^2=1.5$ GeV$^2$.

Finaly we have to define $k'^2_\pm$ and $k_{t\pm}$.
\begin{equation}
\vec k_{t-}=-\vec K_t+z\vec q_t\;\;\;\;\;\;\;
\vec k_{t+}=\vec K_t+(1-z)\vec q_t
\end{equation}
and
\begin{equation}
k'^2_-=\frac{z(1-z)Q^2+m^2_\pi+k^2_{t-}}{z},\;\;\;\;\;\;
k'^2_+=\frac{z(1-z)Q^2+m^2_\pi+k^2_{t+}}{1-z} \;.
\end{equation}
In these notations
$$M^2=\frac{K^2_t+m^2_\pi}{z(1-z)},\;\;\;\;
\;\;\;\; dM^2/dz=(2z-1)\frac{K^2_t+m^2_\pi}{z^2(1-z)^2}$$
and $ z=\frac 12\pm\sqrt{1/4-(K^2_t+m^2_\pi)/M^2}$ with the pion
transverse (with respect to the proton direction) momentum $\vec K_t$
(in the $2\pi$ rest frame) given by expression
$K^2_t=(M^2/4 - m^2_\pi)sin^2\theta$. Note that the positive values of
$cos\theta$ correspond to $z \geq 1/2$ while the negative ones
$cos\theta < 0$ correspond to $z \leq 1/2$.

\subsection{Absorptive correction}

To account for the screening correction we have to consider the diagram
fig. 2, where both pions interact directly with the target. Note that all
the rescatterings of one pion (say $\pi^+$ in fig. 1) are already included
into the $\pi p$ elastic amplitude. The result may be written in form of
eq. (5) with the new factor $\tilde{B}^D$ instead of the old one
$B^D=B^D(\vec K_t,\vec q)$:
\begin{equation}
\tilde{B}^D\; = \; B^D(\vec K_t,\vec q)-\int C\frac{\sigma_{\pi p}
e^{-bl^2_t}} {16\pi^2}B^D(\vec K_t-z\vec l_t,\vec q)d^2l_t
\end{equation}
where the second term is the absorptive correction (fig. 2) and $l_\mu$
is the momentum transfered along the 'Pomeron' loop. The factor $C>1$
reflects the contribution of the enhacement graphs with the diffractive
exitation of the target proton in intermediate state. In accordance with
the HERA data \cite{H2}, where the cross section of "inelastic" (i.e.
with the proton diffracted) $\rho$ photoproduction was estimated as
$\sigma^{inel}\simeq 0.5 \sigma^{el}$ we choose $C=1.5 \pm 0.2$.

\section{$\sigma_L/\sigma_T$ ratio in $\pi^+\pi^-$ electroproduction
near $\rho$-meson peak}

At very large $Q^2$ the background amplitude (5) becomes negligible as,
even without the additional form factor (i.e. at $f(k'^2)\equiv 1$),
the non-resonance cross section falls down as $1/Q^8$ \footnote{In the
amplitude $A_{n.r.}$ one factor $1/Q^2$ comes from the electromagnetic
form factor $F_\pi(Q^2)$ and another one -- from the pion propagator
(term - $z(1-z)Q^2$ in the denominator of $B^D$ (see eq.(6)).}, while
experimentally \cite{zeus1,H11} the $Q^2$ behaviour of the elastic $\rho$
cross sections has been found to be $1/Q^n$, with $n\sim 5 (<8!)$.

Nevertheless, numerically at $Q^2\sim 10\; GeV^2$ the background as well
as interference contributions are still important.

The background as well as interference contributions lead to the
nontrivial behaviour of the ratio $R=\sigma^L/\sigma^T$ with the two pion
mass $M_{2\pi}=M$. In the theoretical formulae which we used the index
$D=L,T$ denotes the polarization of the incoming photon. On the other
hand experimentally one measures the $\rho$-meson polarization, fitting
the angular distribution of decay pions. To reproduce the procedure we
take the flows of initial longitudinal and transverse photons to be equal
to each other ($\epsilon =N^L/N^T=1$, which is close to HERA case) and
reanalyse the sum of cross sections ($\sigma =\sigma^L+\sigma^T$) in a
usual way, selecting the constant and the $cos^2\theta$ parts.
\begin{equation}
I_0\; =\; \int\limits^1_{-1}\sigma(\theta)dcos\theta\; ;\;\;\;\;\;
I_2\; =\; \int\limits^1_{-1}\sigma(\theta)\frac{15}4(3cos^2\theta
-1)dcos\theta
\end{equation}
In these terms the density matrix element $r_{00}=(2I_2+3I_0)/9I_0$ and
\begin{equation}
R\; =\; \sigma^L/\sigma^T\; =\; \frac{r_{00}}{1-r_{00}} =
\frac{3I_0 + 2I_2}{6I_0 - 2I_2}
\end{equation}
Namely these last ratios (12) are presented in fig. 3 for different
$Q^2$ values \footnote{Unfortunately, the results for $R$ presented in
Fig. 6 of our previous paper \cite{RSh} were wrong due to a misprint in
the code.}. We have to recall, that the ratio $R$ given by the last
expression of eq. (12) strictly speaking is not identical to the ratio
of the longitudinal to transverse photon cross sections
$\sigma^L/\sigma^T$. Due to the form factor $f(k'^2)$ the background
pion angular distribution (corresponding to the non-resonant amplitude
(5), (6)) are more complicated than the trivial
$|d^1_{01}(\theta)|^2 = 1/2 \sin^2 \theta $ (for $\sigma^T$) and
$|d^1_{00}(\theta)|^2 = \cos^2 \theta$ (for $\sigma^L$). For example,
even in the case of the transverse photon polarization vector (i.e.
$\sigma^T$) at very large $Q^2$ the non-resonant contribution
$|A_{n.r.}|^2$ reveals the peaks in the forward and backward
($\cos \theta \approx \pm 1$) directions instead of a pure
$\sin^2 \theta$ behaviour. That is why in Fig. 3 we present not the
theoretical $\sigma^L/\sigma^T$ ratio, but the values of
$R = \frac{3I_0 + 2I_2}{6I_0 - 2I_2}$ which are more close to the ratios
measured experimentally. One can see that they depend both on $Q^2$ and
$M$.

Note that for the resonant amplitude $A_{\rho}$ in Fig. 3 we have fixed
$R_{\rho} = const = 2$ independently on the mass $M$ and $Q^2$. On the
other hand for the non-resonant production ($|A_{n.r.}|^2$), neglecting
for simplicity the form factor (i.e. $f(k'^2) = 1$) and the absorptive
corrections, one has $R_{n.r.} = \sigma^L_{n.r.}/\sigma^T_{n.r.}
\approx Q^2/M^2$.

At large $Q^2 \sim$ 10 GeV$^2$ it gives $R_{n.r.} \gg 2$. At last we have
to take into account the interference, which is positive (constructive)
at small $M < M_{\rho}$ and destructive at  $M > M_{\rho}$. Thus the
final curves 3 and 4 fall down with $M$ giving $R \approx 2.5$ at
$M = 0.6$ GeV and  $R \approx 1.5$ at $M = 1.1$ GeV.

Of course at very large $Q^2$ the background amplitude dies out but
simultaneously the value of $R_{n.r.} \approx Q^2/M^2$ increases with
$Q^2$. Therefore the curves 3 and 4 (which correspondent to $Q^2$ =
10 GeV$^2$ and 30 GeV$^2$) are rather close to each other.

In the region of not too large $Q^2 \sim 1$ GeV$^2$ the non-resonant
amplitude is compartible with the resonant one at $M \geq 1$ GeV. In
particular, for $Q^2 = 1$ GeV$^2$ (curve 1) the destructive interference
at $M = 1.2$ GeV cancel the main part of the transverse cross section,
leading to a very large value of $R$ ($\sim 7$). This effect depends
on the $Q^2$ value. Say, the maximum of the peak of $R$ is at 
$Q^2 \approx$  0.03 GeV$^2$ for $M$ = 1.15 GeV and at 
$Q^2 \approx$ 0.3 GeV$^2$ for $M$ = 1.25 GeV. However one must not
take the presented predictions in Fig. 3 too seriously for large $M$.
The interference with the next (heavier) resonances (say, $\rho(1450)$)
was not taken into account in Fig. 3 but at $M > 1.1$ GeV their
contribution can be significant that can change the results.

\section{Conclusion}
We presented the results of $R=\sigma^L/\sigma^T$ ratio calculations
for $\rho$-meson electroproduction with accounting for the background
to 'elastic' mechanism. The role of $\rho$-meson -- background
interference decreases with $Q^2$ however it is not negligible even at
$Q^2\sim 10$ GeV$^2$.

We are grateful to M.Arneodo for stimulating discussions. The paper is
supported by INTAS grant 93-0079.

\newpage

\begin{center}
{\bf Figure captions}\\
\end{center}

Fig. 1. Feynman diagram for the two pion photo-(electro)production.

Fig. 2. Diagram for the absorbtive correction due to both pions
rescattering.

Fig. 3. The ratio $R=\sigma^L/\sigma^T$ in electroproduction process as
a function of pion pair mass at $Q^2 =$ 1 GeV$^2$ (curve 1), 4 GeV$^2$
(curve 2), 10 GeV$^2$ (curve 3) and 30 GeV$^2$ (curves 4) with (solid
curve) and without (dashed curve) form factor.

\begin{figure}
\centerline{\epsfig{file=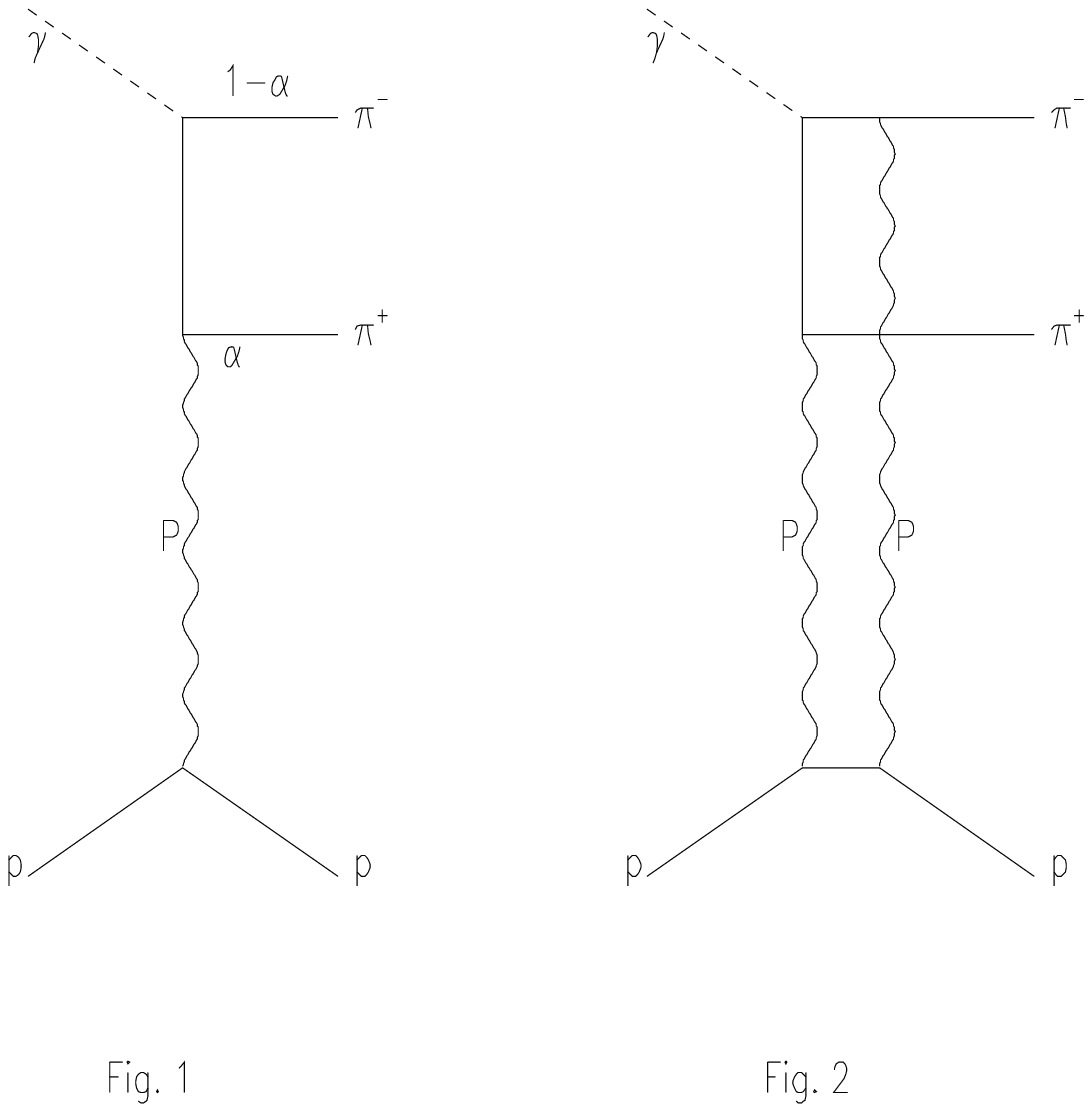,width=16cm}}
\end{figure}

\begin{figure}
\centerline{\epsfig{file=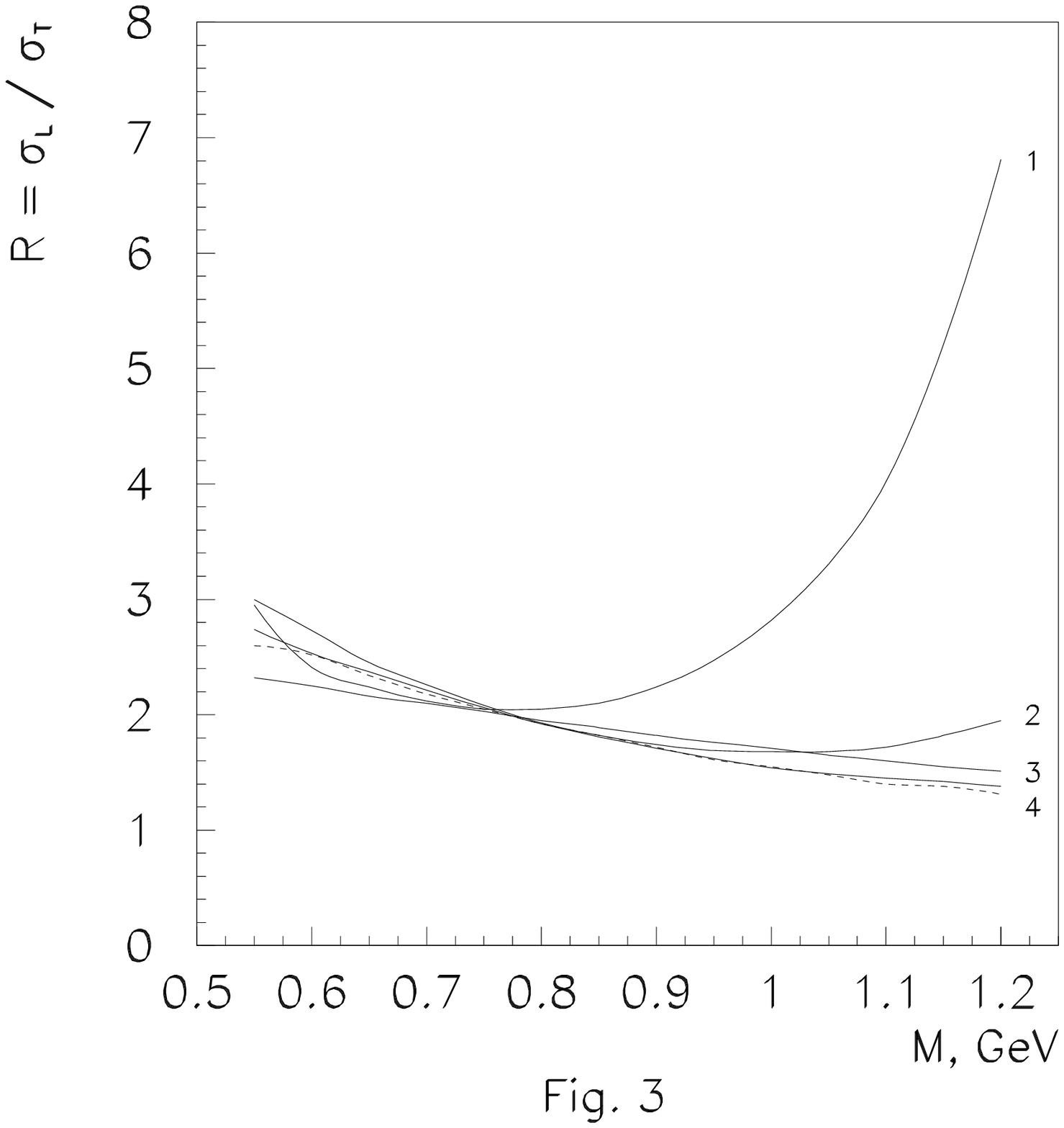,width=16cm}}
\end{figure}

\end{document}